\documentstyle[11pt,aaspp4]{article}

\begin{document}

\title{The Norris Survey of the Corona Borealis Supercluster: \\
III. Structure and Mass of the Supercluster}

\author{Todd A. Small}
\affil{Institute of Astronomy, University of Cambridge, 
Madingley Road, Cambridge CB3 0HA, UK; Palomar Observatory, California
Institute of Technology, Pasadena, CA 91125 \\
Electronic mail: tas@ast.cam.ac.uk}

\author{Chung-Pei Ma}
\affil{Dept.~of Physics and Astronomy,
University of Pennsylvania, Philadelphia, PA 19104 \\
Electronic mail: cpma@dept.physics.upenn.edu}

\author{Wallace L.W. Sargent}
\affil{Palomar Observatory, California Institute of
Technology, Pasadena, CA  91125 \\
Electronic mail: wws@astro.caltech.edu}

\author{and Donald Hamilton}
\affil{Institute of Astronomy and Astrophysics, University of
Munich, Scheinerstrasse 1, D-81679, Munich, Germany \\
Electronic mail: ham@usm.uni-muenchen.de}

\begin{abstract}
 
We present a study of the structure and dynamics of the Corona
Borealis Supercluster ($z \approx 0.07$) based on the redshifts of 528
galaxies in the supercluster.  The galaxy distribution within Corona
Borealis is clumpy and appears overall to be far from relaxed.
Approximately one-third of the supercluster galaxies lie outside of
the Abell clusters in the supercluster.  A background supercluster at
$z \approx 0.11$ makes a substantial contribution to the projected
surface density of galaxies in the Corona Borealis field.  In order to
estimate the mass of the supercluster, we have assumed that the mass
of the supercluster is proportional to $v^2r$, where $v$ and $r$ are
suitable scale velocity and radius, respectively, and we have used
$N$-body simulations of both critical- and low-density universes to
determine the applicability of standard mass estimators based on this
assumption.  Although superclusters are obviously not in equilibrium,
our simulations demonstrate that the virial mass estimator yields mass
estimates with an insignificant bias and a dispersion of only $\sim
25$\% for objects with overdensities $\gtrsim 5$.  Non-uniform spatial
sampling can, however, cause systematic underestimates of as much as
30\%.  The projected mass estimator (Bahcall \& Tremaine 1981) is less
accurate but still provides useful estimates in most cases.  All of
our simulated superclusters turn out to be bound, and based on the
overdensity of the Corona Borealis supercluster, we believe it is also
very likely to be bound and may well have started to collapse.  The
mass of Corona Borealis is at least $3 \times 10^{16} h^{-1} M_{\sun}$
($h$ is the Hubble constant in units of 100 km s$^{-1}$ Mpc$^{-1}$) ,
which yields a $B_{AB}$-band mass-to-light ratio of $564 h {M
\overwithdelims () L}_{\sun}$ on scales of $\sim 20h^{-1}$ Mpc.  The
background supercluster has a similar mass-to-light ratio of $726h {M
\overwithdelims() L}_\odot$.  By comparing the supercluster
mass-to-light ratios with the critical mass-to-light ratio required to
close the universe, we determine that $\Omega_0 \gtrsim 0.4$ on
supercluster scales.

\end{abstract}


\section{Introduction}

Abell (1958), from his survey of galaxy clusters in the Palomar
Observatory Sky Survey, was the first to note the existence of
clusters of clusters of galaxies, which he called ``second order
clusters'' and which have since been dubbed ``superclusters.'' They
are among the largest identified objects in the universe but are only
$\sim 5$ to $\sim 40$ times denser than the field (Small, Sargent, \&
Hamilton 1997a, Paper II in the current series).  In contrast, the
overdensity of an object that has just become virialized is $\sim
200$, and the overdensity in the center of an Abell cluster is $\sim
1000$.  Since the dynamical times of superclusters are comparable to
the Hubble time, superclusters are not relaxed and should therefore
bear the imprints of the physical processes that were dominant during
their formation.  One hopes that studies of superclusters will
ultimately yield information about the nature of density fluctuations.
In addition, since superclusters are mildly or modestly non-linear,
the structure of superclusters may offer clues about the growth of
structure.  In this paper, we describe our efforts to use dynamical
studies of superclusters to provide insights into the distribution of
matter on large scales.

Knowledge of the mass distribution on large scales is, of course,
essential for determining $\Omega_0$, the ratio of the present-day matter
density of the universe to the critical density required for a closed
universe.  On the scales of rich clusters of galaxies ($\sim 1 h^{-1}$
Mpc, where $h$ is the Hubble constant in units of 100 km s$^{-1}$
Mpc$^{-1}$), $\Omega_0$ may be estimated by comparing the
mass-to-light ratio ($M/L$ ratio) of virialized clusters with the
$M/L$ ratio required to close the universe (e.g., Gunn 1978; Kent \&
Gunn 1982; Kent \& Sargent 1983; Sharples, Ellis, \& Gray 1988).  The
comprehensive work by Carlberg et al.\ (1996, 1997) based on 14 rich
clusters yields $M/L = 213 \pm 59 h {M \overwithdelims() L}_\odot$ in
the $r$ band, which in turn gives $\Omega_0 = 0.19 \pm 0.06 \pm 0.04$
(formal 1 $\sigma$ random errors and estimated systematic errors),
well short of the matter density required to close the universe.
However, measurements of $\Omega_0$ on larger scales ($\sim 10 - 100
h^{-1}$ Mpc) from velocity flows and redshift-space distortions tend
to favor larger values, although with large error bars.  (See Dekel,
Burstein, \& White 1996 for a review.)  The dynamics of superclusters
offer an important independent means of estimating $\Omega_0$ on $\sim
20h^{-1}$ Mpc scales.  Here, we present a measurement of the $M/L$
ratio of the Corona Borealis supercluster, and thus a measurement of
$\Omega_0$, based on data from a large redshift survey of the Corona
Borealis supercluster conducted with the 176-fiber Norris Spectrograph
on the Palomar 5m telescope.

The Corona Borealis supercluster is the most prominent example of
superclustering in the northern sky.  Using the ``Lick Counts,'' Shane
\& Wirtanen (1954) were the first to remark on the extraordinary cloud
of galaxies that constitute the supercluster.  Abell also noted the
Corona Borealis supercluster and included it in his catalog of
``second-order clusters.''  Indeed, the Corona Borealis supercluster
includes seven Abell clusters at $z \approx 0.07$ in a 36 deg$^2$
region on the sky and contributes significant power to the two-point
correlation function of nearby Abell clusters (Postman, Geller, \&
Huchra 1986).  In the same region, there are five background Abell
clusters, three of which are at $z \approx 0.11$.  The presence of
this background supercluster at $z \approx 0.11$ was first noted by
Shane \& Wirtanen (1967) using brightest cluster galaxies as a
distance indicator.  Counts of galaxies in the field of the
supercluster, which include the background clusters, show a factor of
3 excess over counts in similarly high galactic latitude fields for
$16\ {\rm mag} \lesssim {\rm Gunn}\ r \lesssim 18\ {\rm mag}$ (Picard
1991).  Kaiser \& Davis (1985) explored whether a structure as large
as the Corona Borealis supercluster is consistent with initially
Gaussian density fluctuations and concluded, with assumptions which
are, in fact, in agreement with our observations, that it is.

The dynamics of the Corona Borealis supercluster have previously been
studied by Postman, Geller, \& Huchra (1988).  They collected 182
redshifts for galaxies in the field of the supercluster, although not
all of these lie in the redshift range of the supercluster.  They
mainly observed galaxies near the cores of the Abell clusters
contained within the supercluster.  By adding up the virial masses of
the Abell clusters, they concluded that the lower limit to the mass of
the supercluster is $2.4 \times 10^{15} h^{-1}$ $M_\odot$.  They also
computed that if the $M/L$ ratio on supercluster scales is comparable
to that on cluster scales, then the supercluster mass is $8.2 \times
10^{15} h^{-1}$ $M_\odot$.  Our aim in this project is to extend the
work of Postman et al.\ (1988) by substantially increasing the number
of redshifts for galaxies in the Corona Borealis supercluster and
therefore more accurately delineate the structure of the supercluster
and obtain a more reliable estimate of its mass.

Due to their large angular size, superclusters have rarely been
studied in detail.  A notable exception is the Shapley supercluster, a
collection of 20 clusters in the Centaurus-Hydra region at $z \sim
0.046$, which has been studied extensively at optical (Quintana et al.\
1995) and X-ray (Ettori, Fabian, \& White 1997) wavelengths .  In
particular, Ettori et al.\  (1997) have used X-ray observations to
study the mass distribution of the Shapley supercluster.  They find
that the core of the supercluster, a region with radius $7.7 h^{-1}$
Mpc centered on Abell 3558, has a mass of $2 - 4 \times 10^{15}$
$M_\odot$ and is likely to be reaching the point of maximum expansion.
There has been no attempt yet, however, to measure the $M/L$ ratio of
the Shapley supercluster.

This paper, the third in the series of papers presenting the results
from the Norris Survey of the Corona Borealis supercluster, is
organized as follows.  In \S\ 2, we give the technical details of the
survey, describe our visual impressions of the Corona Borealis
supercluster, and consider the geometry of the supercluster.  We use
$N$-body simulations of superclusters in \S\ 3 as a test of accurate
techniques for estimating the supercluster mass.  In \S\ 4, we apply
the techniques developed in \S\ 3 to the Corona Borealis supercluster
and to the background supercluster at $z \approx 0.11$.  We discuss
our results, including the value of $\Omega_0$ implied by our
analysis, in \S\ 5.

\section{The Structure of the Corona Borealis Supercluster}

The Norris Survey of the Corona Borealis supercluster has been
described in detail in Small, Sargent, \& Hamilton (1997b), Paper I of
the current series, and will be only briefly reviewed here.  The core
of the supercluster covers a $6\arcdeg \times 6\arcdeg$ region of the
sky centered at right ascension $15^h20^m$, declination $+30\arcdeg$
and consists of seven rich Abell clusters at $z \approx 0.07$.  Since the
field-of-view of the 176-fiber Norris Spectrograph is 20\arcmin\ in
diameter, we planned to observe 36 fields arranged in a rectangular
grid with a grid spacing of 1\arcdeg.  (The precise location of the
fields would be adjusted, typically by 15\arcmin\ and occasionally by
nearly half a degree, in order to maximize the number of fibers on
bright galaxies or to avoid bright stars which had saturated a
significant portion of the field on the original plates.)  We mainly
tried to avoid the cores of the Abell clusters since redshifts for
many galaxies in the cores are available from the literature.  We
successfully observed 23 of the fields and 9 additional fields along
the ridge of galaxies between Abell 2061 and Abell 2067, yielding
redshifts for 1491 extragalactic objects.  We extended our survey with
163 redshifts from the literature, resulting in 1654 redshifts in the
entire survey.  A total of 528 galaxies, 419 with newly-measured redshifts
and 109 from the literature, lie in the redshift range of the
supercluster, $0.06 < z < 0.09$.  (We describe how we chose this
redshift range in \S\ 3 below.)  The velocity errors in our sample are
typically $\sim 75$ km s$^{-1}$.

Figure \ref{figures:cb_on_sky1} shows the surface overdensity of
galaxies with $r \le 19$ mag from our photometric catalog in the
supercluster field.  In order to emphasize the structure of the
supercluster, we have subtracted the mean integrated field galaxy
counts measured in high Galactic latitude fields by Weir, Djorgovski,
\& Fayyad (1995) and then divided by this number to obtain the
projected surface overdensity.  The Abell clusters in the field,
including the ones that are more distant than the supercluster, stand
out prominently.  Four of the Abell clusters (A2056, A2065, A2079, and
A2089) are grouped together in the southern part of the supercluster,
A2061 and A2067 are close together in the northern part, and A2092 is
isolated in the northeastern part.  Only in the diamond-shaped region
delineated by A2056, A2065, A2079, and A2089 can an extended area with
excess galaxy counts be discerned.

In Figure \ref{figures:cone_30000}, we plot redshift-right-ascension
pie diagrams for all galaxies in our survey with $z < 0.15$.  The
diagrams are split by declination as indicated in the figure.  The
Corona Borealis supercluster is sharply delimited along the
line-of-sight by foreground and background underdense regions.  The
well defined boundaries of the supercluster lead us to restrict our
analysis to galaxies with $0.06 < z < 0.09$.  The group of galaxies at
$cz \approx 10000$ km s$^{-1}$ is part of the ``Great Wall'' described
by Geller \& Huchra (1989).  The background supercluster at $z\approx
0.11$ is also evident.

The locations on the sky of all survey galaxies for which we have
obtained redshifts are plotted in Figure \ref{figures:cb_radec}.  The
galaxies marked by large dots have redshifts which place them within
the Corona Borealis supercluster ($0.06 \le z \le 0.09$).  The twelve
large circles, each of which has a radius of roughly $1 h^{-1}$ Mpc,
mark the positions of the Abell clusters projected into the field of
the supercluster; the seven whose names are underlined are the Abell
clusters with redshifts which place them in the supercluster.
Seventeen fields, the six at right ascension 15$^h$13$^m$, the two
southernmost ones at right ascension 15$^h$17$^m$, and all nine along
the A2061-A2067 ridge, were observed when only a 1024$^2$ CCD was
available at Palomar, thus reducing the number of usable fibers by a
factor of two.  The precise positions of the observed fields are
listed in Paper I.  The total number of galaxies for which redshifts
were successfully measured (i.e., including galaxies not in the Corona
Borealis supercluster) ranges from 10 to 42 for the fields observed
with a 1024$^2$ CCD and from 59 to 87 for fields observed with a
2048$^2$ CCD.  Thus, the Norris fields which contain only a few
supercluster galaxies are sparsely populated because the supercluster
is truly not dense in those regions.  The rapid decline in galaxy
density around the A2061-A2067 ridge is particularly striking.  A
similarly complex and irregular distribution of galaxies has been
found in the Shapley supercluster (Quintana et al.\ 1995).

In order to quantify our visual impressions of the Corona Borealis
supercluster, we have computed the fraction of galaxies in the
supercluster which belong to the Abell clusters.  In addition to the
seven catalogued Abell clusters, we also include an additional cluster
which we have identified at R.A. $15^h29^m$, Decl. $+29^\circ 08^m$.
We define a galaxy as belonging to an Abell cluster if its projected
separation from the nearest Abell cluster on the sky is less than $3
h^{-1}$ Mpc (2 Abell radii) and its velocity is less than $3\ \sigma$,
where $\sigma$ is the velocity dispersion, different from the nearest
Abell cluster's mean velocity.  We compute the volume densities for
galaxies in the Corona Borealis supercluster associated and not
associated with the Abell clusters using the methods described in
Paper II.  The Abell clusters occupy 42\% of the supercluster volume,
and the galaxy volume density within the Abell clusters is $0.57 h^3$
Mpc$^{-3}$.  The volume density of galaxies not associated with the
Abell clusters is $0.24 h^3$ Mpc$^{-3}$.  Thus, approximately
two-thirds of the galaxies in the supercluster are associated with one
of the Abell clusters.  About half of the galaxies have projected
separations from the nearest Abell cluster of less than $1.5 h^{-1}$
Mpc (1 Abell radius).

The detailed structure of the supercluster, in particular the fact
that the component not associated with the Abell clusters accounts for
only one-third of the galaxies, is consistent with expectations from
both large redshift surveys and theoretical analyses that the
supercluster is being constructed from infalling clusters which were
formed outside of the supercluster.  The largest published redshift
survey, the Las Campanas Redshift Survey (Shectman et al.\ 1996),
reveals a web-like pattern in which galaxies lie on filaments and
sheets surrounding underdense regions.  Clusters and, more rarely,
superclusters form at the intersections of the filaments (Doroshkevich
et al.\ 1996).  Similar patterns in large-scale structure are revealed
in $N$-body cosmological simulations.  In addition, theoretical
analyses of the merger history of dark matter halos (e.g., Lacey \&
Cole 1993) indicate that a large halo often forms from the merging of
a small number of pieces (which, of course, have themselves formed
from even smaller sub-units).

The true geometry of the supercluster, both in the plane of the sky
and along the line of sight, is difficult to determine.  Bahcall
(1992) has marshalled circumstantial evidence to argue that while the
region containing the seven Abell clusters is only $\sim 20h^{-1}$ Mpc
on a side in the plane of the sky, the entire supercluster extends for
at least $\sim 100h^{-1}$ Mpc.  First, the far side of the Bo\"otes
void, at right ascension $14^h30^m$, declination $+50^\circ$,
(Kirshner et al.\ 1987) is also at a redshift of $z \approx 0.07$.
Second, one of the peaks in the redshift distribution of the
Broadhurst et al.\ (1990) pencil-beam survey of the Galactic Poles is
at the redshift of the Corona Borealis supercluster, even though the
north Galactic Pole is $45^\circ$ away from the core of the
supercluster.  Larger surveys such as the ongoing Center for
Astrophysics Century Survey and the planned Sloan survey will
presumably be able to delineate the true extent of the supercluster on
the plane of the sky.

The core of the supercluster is dramatically elongated along the line
of sight in redshift space.  However, the elongation in redshift space
could in principle be due to a true elongation in real space, to
peculiar velocities, or to a combination of both.\footnote{ A
straightforward approach to distinguishing between the two
possibilities is to compare the apparent magnitude distributions of
samples of supercluster galaxies selected by redshift.  Assuming that
the luminosity function does not vary within the supercluster, samples
of galaxies with lower mean redshifts should appear systematically
brighter than samples with higher mean redshifts (c.f., Mohr, Geller,
\& Wegner 1996).  The strength of this test depends on having a
luminosity function which varies strongly with magnitude.
Unfortunately, the supercluster LF is quite flat over the observed
absolute magnitude range (Paper II), and the constraints which can be
determined by comparing the apparent magnitude distributions of
samples of supercluster galaxies are too weak to be useful.}  We
represent our uncertainty about the true geometry of the supercluster
by the dimensionless parameter $f$, which is the ratio of the
redshift-space to real-space elongation of the supercluster.  The
depth of the supercluster in real space is $\Delta r = (\Delta s/f) h^{-1}$
Mpc, where $\Delta s = c\Delta z/H_0$ is the elongation of the
supercluster in redshift space.  With $\Delta z = 0.03$, $\Delta s =
90 h^{-1}$ Mpc.  If the depth along the line of sight is comparable to
the diameter of the core of the supercluster on the plane of the sky,
$\sim 20h^{-1}$ Mpc, then $f \approx 5$.  In contrast, $f \approx 1$
corresponds to the case in which the peculiar velocities are
negligible and the elongation of the supercluster in redshift space is
similar to the elongation in real space.  The evidence described above
for the $100 h^{-1}$ Mpc extent of the supercluster on the sky,
combined with the theoretical expectation that very large structures
will collapse into pancakes (Zeldovich 1970), leads us to favor the
conclusion that the apparent elongation of the supercluster along the
line of sight is mainly due to peculiar velocities.  In addition, the
large sheets (e.g., the Great Wall, Geller \& Huchra 1989) which
appear to form the ``skeleton'' of the galaxy distribution have widths
of $5-10 h^{-1}$ Mpc (Doroshkevich et al.\ 1996, Dellantonio, Bothun,
\& Geller 1996).  Since the width of the Corona Borealis supercluster
on the plane of the sky is at least $\sim 20h^{-1}$ Mpc, it is
unlikely that Corona Borealis is a sheet aligned along our line of
sight.  To be conservative, the simulated superclusters which we have
used to guide our analysis of the Corona Borealis supercluster (see
below) are chosen to have overdensities comparable to the lowest
possible overdensity of the Corona Borealis supercluster (i.e., when
the supercluster really is elongated in real space, $f \approx 1$).

\section{$N$-body Simulations of Superclusters}

Since superclusters are unrelaxed and contain obvious substructure
(e.g., the Abell clusters themselves), a traditional dynamical
analysis seems of questionable utility.  From dimensional arguments,
however, we expect the mass of the system to be proportional to
$v^2r$, where are $v$ and $r$ are a suitable scale velocity and
radius, respectively.  We aim to test this expectation by analysing
simulated superclusters extracted from large $N$-body cosmological
simulations.  The simulated superclusters, which are in general quite
spatially anisotropic, also enable us to assess the effects of the
non-uniform sampling in our observations on our mass estimates.

As first guesses at successful forms for the mass estimator, we have
chosen the virial mass estimator and the projected mass estimator.  
The standard virial mass estimator is given by
\begin{equation}
\hat{M}_V = {2\sigma_{3D}^2 \over G}
{1 \overwithdelims \langle\rangle r}^{-1}\,,
\end{equation}
where $\sigma_{3D}$ is the 3-dimensional velocity dispersion of the system
and ${\langle 1/r \rangle}^{-1}$ is the mean harmonic radius.
For a bound system (and neglecting projection effects), we in fact expect 
the virial mass estimator never to overestimate the
true mass by more than a factor of 2.
This statement follows by
considering the three cases for the relationship
of the kinetic energy, $K = M_T\, \sigma_{3D}^2/2$, 
and the potential energy,
$W = -GM_T ^2 {\langle 1/r\rangle}/2$, for a bound system
and the implied relationship between the estimated mass
$\hat{M}_V$ and the true mass $M_T$:
\begin{eqnarray}
\label{eq:2kw}
{\rm case \: 1:} & {1 \over 2}|W| < K < |W|\,, & M_T < \hat{M}_V < 2M_T
								 \nonumber\\
{\rm case \: 2:} & K = {1 \over 2}|W|\,, & \hat{M}_V = M_T \\
{\rm case \: 3:} & K < {1 \over 2}|W|\,, & \hat{M}_V < M_T \nonumber
\end{eqnarray}
Thus, the worst the virial estimator can do for a bound system is
to overestimate its mass by a factor of 2, and this occurs only for
a marginally bound system with $K = |W|$.  When $K < |W|/2$,
the virial mass estimator provides a lower limit to the mass of
the system.  In our $N$-body experiments described below,
we find that all 16 candidate superclusters are bound and that
the ratio $\hat{M}_V/M_T$ satisfies the relationship given
in equation~({\ref{eq:2kw}).  We will present evidence in \S\ 4 that
the Corona Borealis supercluster is bound.
 
The virial mass may be calculated from observables using
\begin{equation}
\hat{M}_V = {3\pi \over G} \sigma^2 
{1 \overwithdelims \langle\rangle r_p}^{-1},
\end{equation}
where $\sigma$ is the line-of-sight velocity dispersion and
$\langle 1/r_p \rangle^{-1}$ is the mean harmonic projected separation,
\begin{equation}
{1 \overwithdelims \langle\rangle r_p}^{-1} = {D \over 2}N(N-1)
\biggl(\sum_i \sum_{j < i} {1 \over \theta_{ij}}\biggr)^{-1},
\label{eq:rp}
\end{equation}
where $\theta_{ij}$ is the angular separation of galaxies $i$ and $j$,
$D$ is the radial distance to the cluster, and $N$ is the total number
of galaxies.  This estimator for $\langle 1/r_p \rangle^{-1}$ is very
sensitive to close pairs and is thus quite noisy, especially for
systems which have not been uniformly sampled spatially.
An alternative estimator of
$\langle 1/r_p \rangle^{-1}$ which is less sensitive to irregular
sampling and close pairs has been introduced by Carlberg et al.\ (1996).  Their
``ringwise'' harmonic mean radius is defined by,
\begin{equation}
{1 \overwithdelims \langle\rangle r_p} = {N(N-1) \over 2}
\sum_i \sum_{j < i} {2 \over \pi (r_i + r_j)} K(k_{ij}),
\end{equation}
where $r_i$ and $r_j$ are the projected radii of objects $i$ and $j$,
$K(k)$ is the complete elliptic integral of the first kind in Legendre's
notation (Press et al.\ 1992), and $k_{ij}^2 = 4r_ir_j/(r_i+r_j)^2$.
Although this estimator was originally developed for systems with
circular symmetry on the sky, such as galaxy clusters, we find
(see below) that this estimator does give less biased values of
$\langle 1/r_p \rangle^{-1}$ than the straightforward sum in
equation~(\ref{eq:rp}).

The projected mass estimator (Bahcall \& Tremaine 1981) is given by
\begin{equation}
\hat{M}_P = {f_{\rm PM} \over G N} \sum_i v_{zi}^2 r_{\perp i}\,,
\end{equation}
where $v_z$ is the velocity in the cluster frame and
$r_\perp$ is the projected separation from the cluster
center.
It is designed to give equal weights to particles at all
distances (if $v^2 \propto 1/r$ on the average), but the estimate
depends on the mean eccentricity of the orbits parameterized by
$f_{\rm PM}$.  It can be shown that $f_{\rm PM}=32/\pi$ for isotropic
orbits and $64/\pi$ for radial orbits, independent of the mass
distribution (Heisler, Tremaine,
\& Bahcall 1985).  We have chosen to use $f_{\rm PM} =
32/\pi$ since this yields the smallest masses.  

To test our mass estimators, we have examined 16 simulated
superclusters drawn from $N$-body simulations of structure formation
in both critical- and low-density universes.  The models we chose to
simulate were the standard cold dark matter (CDM) model ($\Omega_0=1$,
$h=0.5$) with a normalization of $\sigma_8=0.7$ for the rms mass
fluctuations in spheres of radius $8\,h^{-1}$ Mpc and a low-density
CDM model with $\Omega_0=0.3$, a cosmological constant
$\Omega_{\Lambda}=0.7$, and $h=0.75$.  The low-density model was
normalized to the 4-year COBE quadrupole $Q_{\rm rms-PS}=18\,\mu K$
(Gorski et al.\ 1996); the corresponding $\sigma_8$ is 0.84.  The
gravitational forces in the simulations were computed with a
particle-particle particle-mesh (P$^3$M) code (Bertschinger \& Gelb
1991).  We have performed both large-box simulations (640 Mpc a side)
with random Gaussian initial conditions and small-box simulations (160
Mpc a side) which were constrained to produce objects with an
overdensity of roughly 5.  The comoving Plummer force softening length
was 160 kpc for all simulations.  The properties of our entire suite
of simulated superclusters are summarized in Table 1, in which we list
the identification number of the simulated supercluster, the
cosmological model, the comoving box size, the total number of
particles in the simulation, the number of particles in the
supercluster, the mass of the supercluster, the volume of the region
containing the supercluster, the overdensity of the supercluster, and
the ratio $2K/|W|$.

To select candidate superclusters from the simulations, we have
searched for clustered regions that consist of multiple dark matter
halos which have not yet merged into one dominant halo.  Overdense
regions which violate this criterion would clearly conflict with the
observations described here and with those of Quintana et al.\ (1995)
of the Shapley supercluster.  In order to test the case in which the
Corona Borealis supercluster is the least dynamically evolved, we have
chosen regions that have an overdensity of $\sim 5$ in a volume of
$30^3$ Mpc$^3$.  This corresponds to the smallest possible density
contrast of the Corona Borealis supercluster, where its elongation in
redshift space is mostly due to physical extension along the line of
sight in real space (i.e., $f \approx 1$).  As we discussed earlier,
however, peculiar velocities are likely to have an important effect,
and the Corona Borealis supercluster is likely to be more compact in
real space than in redshift space ($f>1$) and thus have an overdensity
as high as about 40.  We expect the mass estimators to work better in
this case since the supercluster would then be closer to
virialization, which occurs at an overdensity of $\sim 200$.  Tests of
the mass estimators on smaller regions of our simulated superclusters
with overdensities of $\sim 40$ have verified these expectations.
It is also important to note that the fractions of particles inside
and outside clusters with masses greater than $10^{14} h^{-1}$
$M_\odot$ are typically 2/3 and 1/3, respectively, similar to the
observed fractions of galaxies inside and outside the Abell clusters
within the Corona Borealis supercluster.  In Figure
\ref{figures:sim_sc}, we plot the $x$, $y$, and $z$ projections of
supercluster \#1 to illustrate the type of objects which we have
identified as superclusters in the simulations.

In Figure 5, we plot $\hat M_V/M_T$ (filled squares) and $\hat
M_P/M_T$ (unfilled squares) as a function of $2K/|W|$ for the $x$,
$y$, and $z$ projections of the eight simulated superclusters drawn
from $\Omega_0 = 1$ simulations (panel $a$) and of the eight simulated
superclusters drawn from $\Omega_0 = 0.3$, $\Omega_\Lambda = 0.7$
simulations (panel $b$).  We also plot the data shown in Figure 5 as
histograms in Figure 6 (panels $a$ and $c$).  The correlation between
$\hat M_V/M_T$ and $2K/|W|$ expected from equation~(\ref{eq:2kw}) is
clearly evident in both panels of Figure 5.  For the eight $\Omega_0 =
1$ simulated superclusters, $\langle \hat{M}_V/M_T \rangle = 0.94 \pm
0.24$ and $\langle \hat{M}_P/M_T \rangle = 0.76 \pm 0.24$, where the
averages are over the three projections of the eight simulated
superclusters.  The excellent accuracy of the virial mass estimator
reflects the fact that the mean and dispersion about the mean of
$2K/|W|$ for these eight simulated superclusters are 0.96 and 0.14,
respectively.  We have also examined eight simulated superclusters
drawn from the low-density simulations in order to test the
sensitivities of the mass estimators to cosmology.  Here, too, the
virial mass estimator gives more accurate results than the
projected mass estimator: $\langle \hat M_V/M_T \rangle = 0.99 \pm
0.21$ and $\langle \hat M_P/M_T \rangle = 1.04 \pm 0.52$.  Note,
however, that the projected mass estimator gives large overestimates
of the true mass for superclusters \#14 and \#15, which is due to the
fact that both superclusters have large fractions of their mass at
large radii.  When values of $\hat M_P/M_T$ larger than 1.5 are
excluded, then $\langle \hat M_P/M_T \rangle = 0.78 \pm 0.24$, which
is in line with the result previously obtained for the simulated
superclusters extracted from the $\Omega_0 = 1$ simulations.  We
conclude that the virial mass estimator provides robust estimates of
the masses of superclusters and is insensitive to $\Omega_0$.  
The projected
mass estimator is less accurate and can lead to substantial
overestimates, as much as a factor of two, for superclusters with
unusually large amounts of mass at large radii.

The tests above used all the simulation particles in each
supercluster.  Our observations, however, do not sample every galaxy
in the supercluster.  We have therefore performed simulated
observations of the simulated superclusters to estimate the effects of
irregular sampling in our redshift survey.  For each simulated
supercluster, we projected the supercluster along an axis and chose
only those particles which lie in 16 randomly-positioned fields.  The
total area covered by the 16 fields was 10\% of the total projected
area of the supercluster, matching the fraction of the area of the
Corona Borealis supercluster which we surveyed.  A fraction of the
particles in each field were randomly rejected so that the total
number of particles used in the simulated observations was roughly
500, comparable to the number of galaxies with measured redshifts in
Corona Borealis.  The results of our simulated observations are
summarized in Table 2, in which we record for each supercluster the
masses (relative to the true mass) found by the virial mass estimator
and the projected mass estimator for the $x$, $y$, and $z$
projections.  We also plot the results of our simulated observations
as histograms in Figure \ref{figures:mass_hist} (panels $b$ and $d$).
The non-uniform sampling of the simulated observations tends to reduce
the masses estimated with the virial mass estimator, mainly due to a
systematic underestimate of $\langle 1/r_p \rangle^{-1}$.  The
non-uniform sampling does not significantly affect estimates of the
velocity dispersion.  For the $\Omega_0 = 1$ models, the virial mass
estimator underestimates the true mass by 31\%, and the projected mass
estimator underestimates the true mass by 21\%.  For the low-density
models, the virial mass estimator underestimates the true mass by 5\%,
and the projected mass estimator overestimates the true mass by 27\%.
The dispersion around these values is $\sim 20 - 25$\%.  It is clear
that the ``ringwise'' estimator of the projected mean harmonic radius
is not completely correcting for our non-uniform sampling, which is to
be expected since the estimator is designed for systems with circular
symmetry on the sky.  Nevertheless, the ``ringwise'' estimator does
improve the accuracy of our mass estimates by 20\% relative to mass
estimates using the conventional mean harmonic projected separation
given by equation~(\ref{eq:rp}).  These results indicate that both the
virial mass estimator and the projected mass estimator can be reliably
applied to superclusters, despite the fact that superclusters are
clearly not in equilibrium.  With our irregular sampling, the
estimators generally {\it underestimate} the true mass of the system.
  
Although the circumstantial evidence suggests that the Corona Borealis
supercluster extends for $\sim 100h^{-1}$ Mpc on the sky (\S\ 2), it
is still important to consider the possibility that the Corona
Borealis supercluster has an extreme prolate shape in which we are
looking along the long axis.  Our simulated supercluster \#1 has a
fairly linear geometry with most of the mass concentrated along a
chain of structures.  By observing along this chain, we can assess how
accurately the mass estimators will estimate the true mass in the
pathological case in which the Corona Borealis supercluster is a
cigar-shaped structure pointed directly at us.  For the simulated
supercluster \#1, the virial mass and projected estimators
underestimate the mass of the supercluster by 50\% when the
supercluster is observed along its axis, due, in roughly equal
importance, to reductions in the velocity dispersion and $\langle 1/r_p
\rangle^{-1}$.  The reduction of the velocity dispersion is caused by
coherent infall along the supercluster's axis.  Thus, if the Corona
Borealis supercluster is a prolate structure pointed directly at us,
it is likely we will {\it underestimate} its mass.

\section{The Dynamics and Mass of the Corona Borealis Supercluster}

We demonstrated in the previous section that the virial and projected
mass estimators can be reliably applied to simulated superclusters
that resemble the Corona Borealis supercluster.  Before doing so,
however, we present general considerations of the dynamical
state of the supercluster using the evolution of a spherical
overdensity in an expanding universe as our model (Peebles 1980).  
The supercluster is certainly not perfectly spherical, and our
knowledge of the structure along the line of sight is particularly
unconstrained (\S 2).  However, the models of Eisenstein \& Loeb (1995)
illustrate that the turn-around times of the short axes of a triaxial
perturbation are fairly well predicted by the turn-around time of
a spherical perturbation with the same initial density and that the
overdensity inside the triaxial perturbation is well described
by the spherical model to overdensities of $\sim 50$. 
  
The galaxy number overdensity of the supercluster for $M(B_{AB}) \le
-16.3 + 5 \log h$ mag is $\delta_{SC} = (n_{SC}/\bar n) - 1 \approx
7f$, where $n_{SC}$ and $\bar n$ are the mean number densities of
galaxies in the supercluster and in the field, respectively, and $f$,
introduced in \S 2, is the ratio of the redshift-space to real-space
elongation along the line of sight.  An outer shell of a density
perturbation is bound if the mean overdensity within the shell is
greater than $(\Omega_0^{-1} - 1)/(1+z)$.  There is only a very
small range of positive density perturbations that are not bound
(i.e., $0 < \delta < (\Omega_0^{-1}-1)/(1+z)$), and it is thus
unlikely to observe a supercluster with an appreciable density
contrast which is freely expanding.  Taking $\Omega_0 = 0.3$ as a very
rough lower limit on the density parameter, the overdensity of the
supercluster must be greater than 2.1 for the supercluster to be
bound.  The supercluster is, then, clearly bound.  While bound, the
supercluster could still be expanding and yet to turn around.  The
overdensity required for turn-around has been computed for $\Omega_0 \le
1$ by Silk (1977) and by Reg\H{o}s \& Geller (1989).  For $\Omega_0 =
1$, the turn-around overdensity is the well-known value $(9\pi^2/16) -
1 \approx 4.55$.  For $\Omega_0 = 0.3$, the turn-around overdensity is
$\approx 12$.  From Figure 4 of Reg\H{o}s \& Geller (1989), we see
that the supercluster is likely to be collapsing unless the
supercluster is quite elongated in real space ($f \lesssim 2$) and
$\Omega_0 \lesssim 0.4$.

Before we compute the mass of the whole supercluster, we recompute the
lower limit to the mass of the supercluster determined by simply
adding up the masses of the individual Abell clusters.  For each of
the eight clusters (the seven catalogued clusters plus the cluster
which we have identified at R.A.  $15^h29.2^m$, Decl. $+29^\circ08^m$),
we identify all galaxies within a projected distance of $1.5 h^{-1}$
Mpc (1 Abell radius) of the cluster center and compute the cluster
mass with the virial mass estimator (using the ringwise harmonic mean
projected radius).  We do not have redshifts for enough galaxies in
Abell 2056 to estimate reliably its mass.  The other clusters are,
however, well sampled, ranging from 26 redshifts in Abell 2079 to 105
in Abell 2061.  We use the biweight estimators and bias-corrected,
bootstrap errors for our determinations of the centroid velocities and
dispersions and their associated errors, as recommended by Beers,
Flynn, \& Gebhardt (1990).  The velocity dispersions are corrected to
the cluster rest frames.  The results for the individual clusters are
summarized in Table 3, where we record for each cluster its centroid
velocity, dispersion, harmonic mean projected radius, and virial mass.
The quoted errors are 90\% confidence intervals.  The sum of the
masses of the seven clusters for which we have adequate data is $5.3
\times 10^{15}h^{-1} M_\odot$, a factor of 2 larger than the sum
computed by Postman et al. (1988).  This difference is mainly due to
the fact that we computed the masses within a projected radius of $1.5
h^{-1}$ Mpc, whereas Postman et al.\ (1988) used $1.0 h^{-1}$ Mpc.
Our sum is also based on seven rather than six clusters.

We have plotted the line-of-sight velocity histogram (in the
supercluster frame) for the Corona Borealis supercluster in Figure
\ref{figures:cor_bor_z}.  Again using the techniques of Beers, Flynn,
\& Gebhardt (1990), we estimate the centroid velocity of the Corona
Borealis supercluster to be $c\bar z = 22420^{+149}_{-138}$ km
s$^{-1}$ and the dispersion to be $\sigma = 1929^{+81}_{-67}$ km
s$^{-1}$ in the cluster rest frame.  The errors bars are 90\%
confidence intervals.  Due to our irregular sampling of the
supercluster, which itself has an irregular density distribution, it
is not straightforward to estimate the harmonic mean projected radius.
We have taken two approaches.  First, we have used the ``ringwise''
estimator described by Carlberg et al.\ (1996), which yields $\langle
1/r_p \rangle^{-1} = 4.6 h^{-1}$ Mpc.  Second, we have estimated
$\langle 1/r_p \rangle^{-1}$ using all the galaxies in the supercluster
field with $16\ {\rm mag} \le r \le 19\ {\rm mag}$, whether or not they
have measured redshifts, weighted by the observed redshift
distribution as a function of $r$ magnitude.  This method yields
$\langle 1/r_p \rangle^{-1} = 4.1 h^{-1}$ Mpc, 11\% smaller than the
value obtained using the ``ringwise'' estimator.  The virial mass
estimator gives a mass for the Corona Borealis supercluster of $3.8
\times 10^{16} h^{-1}$ $M_\odot$ using $\langle 1/r_p \rangle^{-1} = 4.6
h^{-1}$ Mpc and $3.3 \times 10^{16} h^{-1}$ $M_\odot$ using $\langle
1/r_p \rangle^{-1} = 4.1 h^{-1}$ Mpc.  The projected mass estimator
yields a similar value, $4.2 \times 10^{16} h^{-1}$ $M_\odot$.  If we
apply the two mass estimators to the eight clusters, treated as test
particles, we also obtain values for the total mass of the
supercluster of $4 \times 10^{16} h^{-1}$ $M_\odot$.  While case 3
of equation~(\ref{eq:2kw}) demonstrates that the virial theorem can
in principle underestimate the mass of a bound system by an arbitrarily
large amount, the results of our simulations, which are recorded
in Table 2 and depicted in Figure~\ref{figures:mass_hist}, indicate
that we are unlikely to underestimate the mass of the Corona Borealis
supercluster by more than a factor of 2.  We thus place a rough
upper limit on the mass of the supercluster of $8 \times 10^{16} h^{-1}$
$M_\odot$, comparable to the upper limit derived by Postman et al.\ (1988).

Given the results of our tests of the virial mass estimator and the
projected mass estimator on simulated superclusters described in \S\
3, we believe that a secure lower bound to the mass of the Corona
Borealis supercluster is $3 \times 10^{16} h^{-1}$ $M_\odot$.  By
integrating the supercluster luminosity function, we can compute the
mean luminosity density of the supercluster and thereby measure the
$M/L$ ratio of the supercluster on scales of $\sim 20 h^{-1}$ Mpc.
The supercluster luminosity function for $M(B_{AB}) \le -16.3 + 5 \log
h$ mag, where we are using the AB-normalized $B$ band (Oke 1974), is
presented in Paper II; a straightforward integration of the luminosity
function yields a mean luminosity density in the supercluster in the
$B_{AB}$ band of $\rho_L(B_{AB}) = 1.9 \times 10^9
h L_\odot$Mpc$^{-3}$.  Taking the solid angle of the survey to be
0.0076 sr (= 25 deg$^2$) and the limits of the supercluster to be at
$z = 0.06$ and $z = 0.09$, the volume of the region surveyed is $2.8
\times 10^4 h^{-3}$ Mpc$^3$.  The $M/L$ ratio of the supercluster in
the $B_{AB}$ band is thus $564h {M \overwithdelims () L}_{\sun}$.  Our
mass estimates for the Corona Borealis supercluster are comparable to
the mass estimates derived by Postman et al.\ (1988) under the
assumption that the differences in the mean redshifts of the
constituent clusters are due to peculiar motions generated by the
supercluster.  Their best estimate of the supercluster mass is,
however, five times smaller than ours because they assumed that the
$M/L$ ratio of the Corona Borealis supercluster was similar to that of
rich clusters and they used a supercluster volume only one-third the
size of the volume we used.

We can repeat our analysis of the Corona Borealis supercluster on the
background A2069 supercluster.  From inspection of Figure 2, we take
the redshift limits of the A2069 supercluster to be $z = 0.10$ and $z
= 0.13$, which gives us 352 galaxies with measured redshifts in the
supercluster.  The galaxy number overdensity of the A2069 supercluster
between these redshift limits and for $M(B_{AB}) \le -17.5 + 5\log h$ mag
is $\delta_{SC} \approx 4f_{A2069}$, where we again parameterize the
depth of the supercluster along the line of sight in real space as
$\Delta r = (\Delta s/f_{A2069})$ Mpc, with $\Delta s = c\Delta z/H_0
= 90 h^{-1}$ Mpc for $\Delta z = 0.03$.  The minimum overdensity is
obtained for $f_{A2069} \approx 1$, which corresponds to the case in
which the peculiar velocities are negligible and the elongation in
redshift space is similar to the elongation in real space.  If, on the
other hand, the depth in real space is similar to the diameter of the
supercluster in the plane of the sky, then $f_{A2026} \approx 5$.  For
$\Omega_0 = 0.3$ and $z = 0.115$, a spherical perturbation with an
overdensity of $(\Omega_0^{-1} - 1)/(1+z) \approx 2.1$ is bound.
Thus, the A2069 supercluster is likely to bound, although possibly
only marginally so if $f_{A2069} \approx 1$.  If the supercluster is,
indeed, only marginally bound, then we would expect, in light of the
discussion in \S 3, that the virial mass estimator may overestimate
the mass of the supercluster, but by no more than a factor of 2.

The centroid velocity and velocity dispersion of the A2069
supercluster are $34338^{+148}_{-140}$ km s$^{-1}$ and
$1684^{+145}_{-151}$ km s$^{-1}$, respectively.  We have plotted the
line-of-sight velocity histogram (in the supercluster frame) of the
A2069 supercluster in Figure \ref{figures:a2069_z}.  As for our
analysis of the Corona Borealis supercluster, we take two approaches
to estimating the mean harmonic projected radius.  The Carlberg et
al. (1996) ``ringwise'' estimator yields $\langle 1/r_p \rangle^{-1} =
6.5 h^{-1}$ Mpc.  We have also estimated the mean harmonic projected
radius using all galaxies with $16 \le r \le 19$ in the supercluster
field, corrected by the measured redshift distribution as a function
of magnitude.  This procedure gives $\langle 1/r_p \rangle^{-1} = 5.2
h^{-1}$ Mpc, a 20\% reduction from the value obtained with the
``ringwise'' estimator.  Using the virial mass estimator, the mass of
the A2069 supercluster is $4.1 \times 10^{16} h^{-1}$ $M_\odot$ with
the larger value of $\langle 1/r_p \rangle^{-1}$ or $3.3 \times
10^{16} h^{-1}$ $M_\odot$ with the smaller value.  With the projected
mass estimator, the mass of the A2069 supercluster is $6 \times
10^{16} h^{-1} M_\odot$.  A firm lower limit to the mass of the A2069
supercluster is, therefore, $3 \times 10^{16} h^{-1}$ $M_\odot$.
Integrating the A2069 supercluster luminosity function, we find that
the luminosity density of the A2069 supercluster is $\rho_L(B_{AB}) =
7.0 \times 10^8 h L_\odot$Mpc$^{-3}$.  We surveyed $5.9 \times 10^4
h^{-3}$ Mpc$^3$ in the A2069 supercluster, and we thus find that the
$M/L$ ratio of the A2069 supercluster in the $B_{AB}$ band, using the
lower limit mass of $3 \times 10^{16} h^{-1}$ $M_\odot$, is $726 h {M
\overwithdelims () L}_\odot$, roughly 30\% higher that the $M/L$ ratio
of the Corona Borealis supercluster.  10\% of the difference is due to
the brighter integration limit for the luminosity density of the A2069
supercluster.  The remaining 20\% difference between the two
superclusters is not, however, likely to be significant, given both
the $\sim 25$\% error in the mass estimators and the possibility that
the mass estimators will tend to overestimate modestly the true mass
of a weakly bound system such as the A2069 supercluster.

\section{Discussion}

As an immediate application of our measurement of the $M/L$ ratio of
the Corona Borealis and A2069 superclusters, we can estimate
$\Omega_0$ on a scale of $\sim 20h^{-1}$ Mpc by comparing the $M/L$
ratio of the superclusters to the $M/L$ ratio required to close the
universe.  As computed from our determination of the local luminosity
function (Paper II), the $B_{AB}$-band luminosity density for galaxies
with $M(B_{AB}) \le -16.3 + 5 \log h$ mag is $1.8 \times 10^8 L_\odot
h$ Mpc$^{-3}$.  The uncertainty in this number is dominated by
systematic errors, and so the most straightforward means to assess the
uncertainty is simply to compare values obtained in independent
redshift surveys of local galaxies.  The largest local survey is the
Las Campanas Redshift Survey (Shectman et al.\ 1996).  The luminosity
function from this survey has been computed by Lin et al. (1996) and
yields a luminosity density in the $B_{AB}$-band of $1.6 \times 10^8
L_\odot h$ Mpc$^{-3}$, where we have used $\langle B_{AB} - r_{\rm
LCRS} \rangle = 0.72$ mag to convert from isophotal hybrid $r_{\rm
LCRS}$ magnitudes to total $B_{AB}$ magnitudes (see Paper II) and we
have assumed a flat slope for the low luminosity end of the luminosity
function.  The luminosity density for $M(B_{AB}) < -16 + 5 \log h$ mag
reported by Zucca et al.\ (1997) for the ESO Slice Project is $2.0
\times 10^8 L_\odot h$ Mpc$^{-3}$.  Finally, the luminosity densities
obtained by integrating the local luminosity functions measured by
Ellis et al.\ (1996) from the Autofib Survey and Ratcliffe et al.\
(1997) from the Durham/UKST Galaxy Redshift Survey are $1.9 \times
10^8 L_\odot h$ Mpc$^{-3}$ and $1.7 \times 10^8 L_\odot h$ Mpc$^{-3}$
in the $B_{AB}$ band, respectively.  Thus, we conclude that the local
$B_{AB}$-band luminosity density is $(1.8 \pm 0.2) \times 10^8 L_\odot
h$ Mpc$^{-3}$ for $M(B_{AB}) \lesssim -16 + 5 \log h$ mag.  The
corresponding critical $M/L$ ratio to close the universe is $1550 \pm
170 h {M \overwithdelims() L}_\odot$.

The final step before deriving a value for $\Omega_0$ is to consider
whether there has been differential luminosity evolution between
galaxies in the superclusters and in the field.  There is no reason,
in principle, why the luminosity per unit mass in the superclusters
should be identical to that in the field.  However, Carlberg et al.\
(1997) report that galaxies in the cores of rich clusters have only
faded by 0.11 mag relative to galaxies in the field, and so we expect
a very small variation between the galaxies in the superclusters,
which are on average significantly less dense than rich clusters, and
the galaxies in the field.  In Paper II, we compared the luminosity
functions of the Corona Borealis and A2069 superclusters to the local
field luminosity function.  We found that the characteristic
luminosity, $L^\ast$, was $\sim 0.5$ mag brighter in the Corona
Borealis supercluster than in the field, while $L^\ast$ in
the A2069 supercluster was very close to the value in the field.
Since the measured value of $L^\ast$ is strongly correlated with the
poorly-constrainted faint end slope of the luminosity function, the
errors on $L^\ast$ are quite large ($\gtrsim 0.3$ mag).  Nevertheless,
there is no evidence for fading of supercluster galaxies relative to
field galaxies; any correction for brightening of supercluster
galaxies relative to field galaxies would raise our estimate of
$\Omega_0$.

The $M/L$ ratio of the Corona Borealis supercluster is $564 h {M
\overwithdelims() L}_\odot$, and therefore we determine that $\Omega_0
= 0.36$ on supercluster scales, or roughly twice the value computed by
Carlberg et al.\ (1996) for rich clusters of galaxies.  Repeating the
above analysis for the A2069 supercluster, but only computing the
luminosity densities for $M(B_{AB}) \le -17.5 + 5 \log h$ mag, gives
$\Omega_0 = 0.44$.  Since our simulated observations of our simulated
superclusters indicate that our mass estimators are likely to
underestimate (by $\lesssim 30\%$) the true masses of the
superclusters, we conclude that $\Omega_0 \gtrsim 0.4$ on $\sim
20h^{-1}$ Mpc scales, which is comparable to estimates of $\Omega_0$
on similar and larger scales based on analyses of large-scale velocity
flows (Strauss \& Willick 1995) and agrees with the ``tentative
consensus'' value reached by Dekel et al.\ (1996).

The principal weakness of our measurement of $\Omega_0$ is, of course,
that it is based on only two superclusters.  However, future
large-area redshift surveys should generate data for a much larger
number of superclusters.  In particular, the imminent 2dF (Colless
1997) and Sloan surveys will densely map substantial fractions of the
sky and provide large numbers of redshifts for many superclusters.

\acknowledgements 

We are grateful to the Kenneth T. and Eileen L. Norris Foundation for
their generous grant for construction of the Norris Spectrograph.  We
thank the staff of the Palomar Observatory for the expert assistance
we have received during the course of the survey, Jim Frederic and
Paul Bode for aid with the $N$-body simulations, and Roger Blandford, David
Buote, and the referee, Marc Postman, for helpful comments.  The
supercomputing time was provided by the National Scalable Cluster
Project at the University of Pennsylvania, the National Center for
Supercomputing Applications, and the Cornell National Supercomputer
Facility.  This work has been supported by an NSF Graduate Fellowship
(TAS), a Caltech PMA Division Fellowship (CPM), and NSF grant
AST92-213165 (WLWS).

\newpage

\begin{deluxetable}{ccccccccc}
\tablecolumns{9}
\tablewidth{0pt}
\tablenum{1}
\tablecaption{Properties of the Simulated Superclusters}
\tablehead{
\colhead{ID \#} &
\colhead{Model\tablenotemark{a}} &
\colhead{Box\tablenotemark{b}} &
\colhead{$N_{tot}$} &
\colhead{$N_{SC}$} &
\colhead{M$_{SC}$} &
\colhead{V$_{SC}$\tablenotemark{b}} &
\colhead{${\delta \rho \over \rho}$} &
\colhead{$2K/|W|$} \\[0.2ex]
\colhead{} &
\colhead{} &
\colhead{($h^{-1}$ Mpc)} &
\colhead{} &
\colhead{} &
\colhead{($10^{15} h^{-1}$ $M_\odot$)} &
\colhead{($10^3 h^{-3}$ Mpc$^3$)} &
\colhead{} &
\colhead{}}
\startdata
1  & CDM 	  & 320 & $256^3$ & 9780  & 5.3 &  3.4 & 4.7 & 0.98 \nl
2  & CDM	  & 320 & $256^3$ & 9657  & 5.3 &  3.4 & 4.6 & 1.07 \nl
3  & CDM	  & 320 & $256^3$ & 10089 & 5.5 &  3.7 & 4.4 & 0.89 \nl
4  & CDM	  & 320 & $256^3$ & 9684  & 5.3 &  3.7 & 4.1 & 0.84 \nl
5  & CDM	  &  80 & $64^3$  & 9694  & 5.3 &  3.4 & 4.6 & 0.85 \nl
6  & CDM	  &  80 & $64^3$  & 13890 & 7.6 &  3.4 & 7.0 & 1.23 \nl
7  & CDM	  &  80 & $64^3$  & 10755 & 5.9 &  3.4 & 5.2 & 1.06 \nl
8  & CDM	  &  80 & $64^3$  & 8619  & 4.7 &  3.4 & 4.0 & 0.79 \nl
9  & $\Lambda$CDM & 480 & $128^3$ & 1265  & 5.6 & 11.4 & 4.9 & 0.98 \nl
10 & $\Lambda$CDM & 480 & $128^3$ & 1261  & 5.6 & 11.4 & 4.8 & 1.08 \nl
11 & $\Lambda$CDM & 480 & $128^3$ & 1390  & 6.1 & 12.3 & 4.9 & 0.85 \nl
12 & $\Lambda$CDM & 480 & $128^3$ & 1261  & 5.6 & 12.5 & 4.3 & 0.83 \nl
13 & $\Lambda$CDM & 120 & $64^3$  & 8251  & 4.6 & 11.4 & 3.8 & 0.84 \nl
14 & $\Lambda$CDM & 120 & $64^3$  & 8631  & 4.8 & 11.2 & 4.1 & 1.23 \nl
15 & $\Lambda$CDM & 120 & $64^3$  & 11477 & 6.3 & 11.4 & 5.6 & 1.20 \nl
16 & $\Lambda$CDM & 120 & $64^3$  & 11483 & 6.3 & 11.4 & 5.6 & 1.07 \nl
\enddata
\tablenotetext{a}{CDM: $\Omega_0 = 1$, $h = 0.5$; $\Lambda$CDM: $\Omega_0
= 0.3$, $\Omega_\Lambda = 0.7$, $h = 0.75$}
\tablenotetext{b}{Comoving.}
\end{deluxetable}

\newpage

\begin{deluxetable}{ccccccc}
\tablecolumns{7}
\tablewidth{0pt}
\tablenum{2}
\tablecaption{Mass Estimates for the Simulated Superclusters from
              Simulated Observations}
\tablehead{
\colhead{ID \#} &
\multicolumn{2}{c}{x-axis} &
\multicolumn{2}{c}{y-axis} &
\multicolumn{2}{c}{z-axis} \\[.2ex]
\colhead{} &
\colhead{${\hat{M}_V \over M_T}$} &
\colhead{${\hat{M}_P \over M_T}$} &
\colhead{${\hat{M}_V \over M_T}$} &
\colhead{${\hat{M}_P \over M_T}$} &
\colhead{${\hat{M}_V \over M_T}$} &
\colhead{${\hat{M}_P \over M_T}$}}
\startdata
1 & $0.61 \pm 0.20$ & $0.76 \pm 0.23$ & $0.59 \pm 0.18$ & $0.66 \pm 0.19$
  & $0.99 \pm 0.36$ & $1.13 \pm 0.41$ \nl
2 & $0.67 \pm 0.21$ & $0.74 \pm 0.19$ & $0.64 \pm 0.21$ & $0.63 \pm 0.18$
  & $0.81 \pm 0.25$ & $0.84 \pm 0.25$ \nl
3 & $0.82 \pm 0.24$ & $0.87 \pm 0.23$ & $0.55 \pm 0.17$ & $0.59 \pm 0.17$
  & $0.55 \pm 0.25$ & $0.64 \pm 0.26$ \nl
4 & $0.71 \pm 0.20$ & $0.84 \pm 0.25$ & $0.65 \pm 0.25$ & $0.83 \pm 0.35$
  & $0.45 \pm 0.12$ & $0.56 \pm 0.16$ \nl
5 & $0.78 \pm 0.26$ & $1.07 \pm 0.33$ & $0.67 \pm 0.26$ & $0.87 \pm 0.36$
  & $0.87 \pm 0.30$ & $1.09 \pm 0.41$ \nl
6 & $1.15 \pm 0.48$ & $1.40 \pm 0.58$ & $0.60 \pm 0.19$ & $0.71 \pm 0.25$
  & $0.59 \pm 0.20$ & $0.66 \pm 0.24$ \nl
7 & $0.63 \pm 0.16$ & $0.65 \pm 0.14$ & $0.52 \pm 0.17$ & $0.51 \pm 0.14$
  & $0.87 \pm 0.27$ & $0.93 \pm 0.24$ \nl
8 & $0.42 \pm 0.13$ & $0.45 \pm 0.14$ & $0.67 \pm 0.13$ & $0.76 \pm 0.17$
  & $0.68 \pm 0.24$ & $0.76 \pm 0.25$ \nl
13& $0.54 \pm 0.29$ & $0.73 \pm 0.39$ & $0.82 \pm 0.27$ & $1.01 \pm 0.35$
  & $0.79 \pm 0.34$ & $1.12 \pm 0.45$ \nl
14& $1.23 \pm 0.46$ & $1.65 \pm 0.58$ & $1.42 \pm 0.55$ & $1.96 \pm 0.73$
  & $1.00 \pm 0.36$ & $1.34 \pm 0.47$ \nl
15& $1.12 \pm 0.58$ & $1.55 \pm 0.82$ & $0.91 \pm 0.40$ & $1.29 \pm 0.51$
  & $1.09 \pm 0.54$ & $1.41 \pm 0.73$ \nl
16& $0.79 \pm 0.22$ & $1.00 \pm 0.28$ & $0.67 \pm 0.36$ & $0.86 \pm 0.43$
  & $1.06 \pm 0.31$ & $1.33 \pm 0.37$ \nl
\enddata
\tablecomments{Because superclusters \#9, \#10,
\#11, and \#12 were drawn from a simulation with a large box size
($640^3$ Mpc$^3$) but with a comparatively small number of particles
($128^3$), simulated observations of 10\% of the area yield only $\sim
120$ particles, and so we have excluded these superclusters from our
analysis of simulated observations.}
\end{deluxetable}

\newpage

\begin{deluxetable}{cccccc}
\tablecolumns{6}
\tablewidth{0pt}
\tablenum{3}
\tablecaption{Galaxy Clusters in the Corona Borealis Supercluster}
\tablehead{
\colhead{Cluster} &
\colhead{$N_z$\tablenotemark{a}} &
\colhead{Centroid} &
\colhead{Dispersion} &
\colhead{${1 \overwithdelims \langle\rangle r_p}^{-1}$} &
\colhead{$M_V$\tablenotemark{b}} \\[.2ex]
\colhead{} &
\colhead{} &
\colhead{km s$^{-1}$} &
\colhead{km s$^{-1}$} &
\colhead{$h^{-1}$ Mpc} &
\colhead{$10^{14} h^{-1}$ $M_\odot$}}
\startdata
A2056 & 10 & \nodata & \nodata & \nodata & \nodata \nl
A2061 & 105 & $23512^{+170}_{-143}$ & $1020^{+180}_{-182}$ & 0.40 &
	$9.08^{+3.49}_{-2.95}$ \nl
A2065 & 31  & $21767^{+369}_{-367}$ & $1203^{+371}_{-289}$ & 0.58 &
	$18.5^{+13.2}_{-8.2}$ \nl
A2067 & 55  & $22435^{+280}_{-220}$ & $ 953^{+175}_{-250}$ & 0.56 &
	$11.1^{+4.5}_{-5.1}$ \nl
A2079 & 26\tablenotemark{c}  & $19656^{+209}_{-239}$ & $ 652^{+295}_{-172}$ & 0.62 &
	$5.82^{+6.48}_{-2.67}$ \nl
A2089 & 30  & $21968^{+136}_{-185}$ & $ 545^{+246}_{-207}$ & 0.73 &
	$4.77^{+5.28}_{-2.94}$ \nl
A2092 & 44  & $20064^{+137}_{-124}$ & $ 581^{+252}_{-138}$ & 0.27 &
	$1.99^{+2.10}_{-0.83}$ \nl
Cl1529+29 & 43\tablenotemark{d} & $25145^{+162}_{-195}$ & $534^{+92}_{-80}$ & 0.29 &
	$1.82^{+0.68}_{-0.50}$ \nl
\enddata
\tablenotetext{a}{Number of galaxies with redshifts with projected
separations $\le 1.5h^{-1}$ Mpc.}
\tablenotetext{b}{Mass estimated with the virial mass estimator.}
\tablenotetext{c}{Two galaxies with $cz > 24000$ km s$^{-1}$ have been
excluded from the dynamical analysis.}
\tablenotetext{d}{Eight galaxies with $cz < 22000$ km s$^{-1}$ have been
excluded from the dynamical analysis.}
\end{deluxetable}

\newpage

\begin{figure}
\plotfiddle{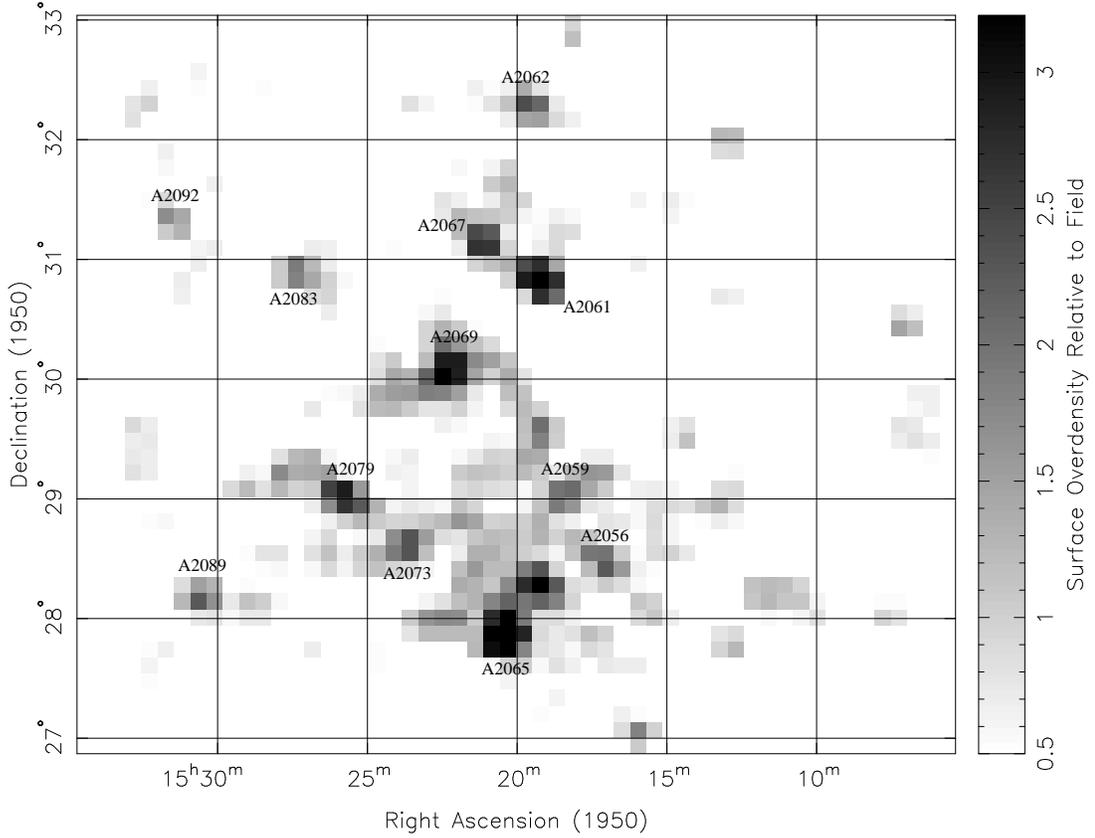}{4.3 in}{-90}{62}{62}{-234}{360}
\caption[]
{A grayscale plot of the galaxy surface overdensity in the field of
the Corona Borealis supercluster for all galaxies brighter than $r =
19$ mag in our photometric catalog.  The galaxy surface overdensity is
the surface density in the supercluster region with the integrated
high Galactic latitude counts (to $r = 19$ mag) from Weir et
al. (1995) subtracted off and then normalized by the subtracted value.
The bar on the right side shows the intensity scale.  While the
Abell clusters stand out prominently, the smoothly distributed
component of galaxies in the supercluster is quite weak and may be
easily discerned only in the diamond-shaped region defined by A2056,
A2065, A2079, A2089.}
\label{figures:cb_on_sky1}
\end{figure}

\begin{figure}
\plotfiddle{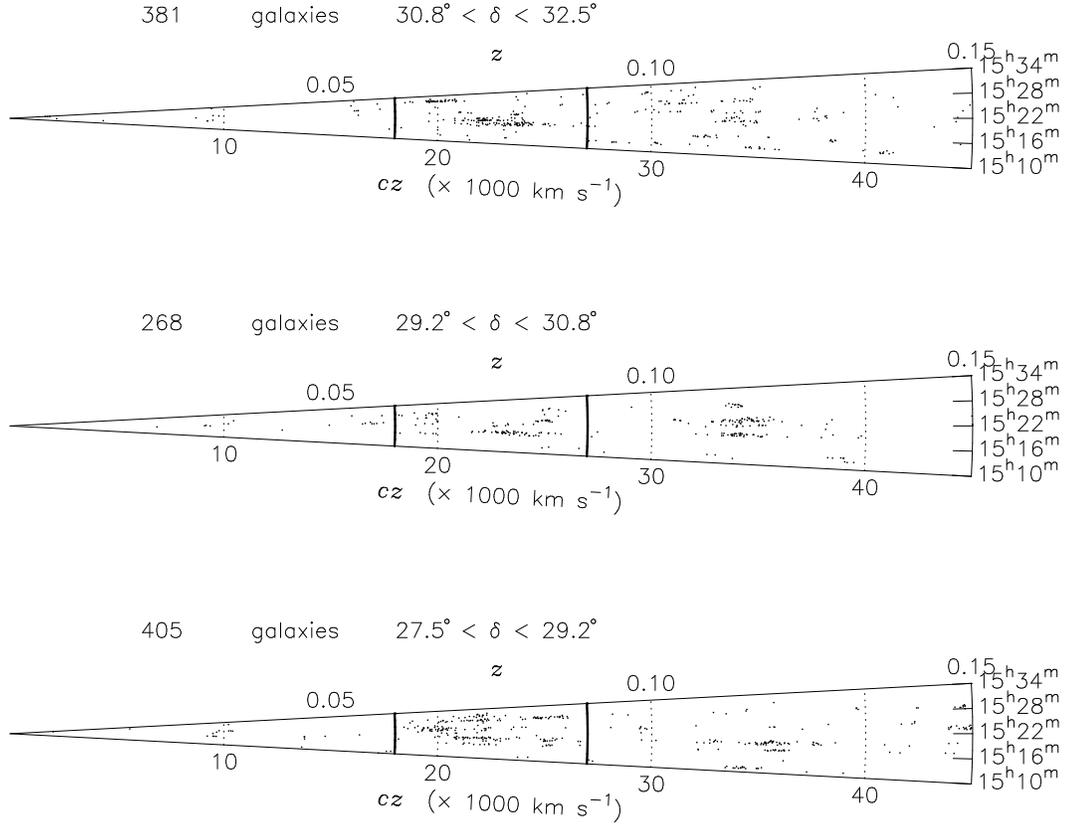}{4.3 in}{-90}{62}{62}{-234}{360}
\caption[]
{Redshift-right-ascension cone diagrams for galaxies in our survey
with $cz < 45000$ km s$^{-1}$, divided into 3 declination slices.  The
Corona Borealis supercluster is the prominent overdense region between
$cz \approx 18000$ km s$^{-1}$ and $cz \approx 27000$ km s$^{-1}$;
these limits are marked in the cone diagrams by the heavy lines.  
The dashed lines mark the background supercluster at $cz\approx 34000$
km s$^{-1}$.  The smaller structure at $cz \approx 10000$ km s$^{-1}$ 
is part of the ``Great Wall'' of galaxies.}
\label{figures:cone_30000}
\end{figure}

\begin{figure}
\plotfiddle{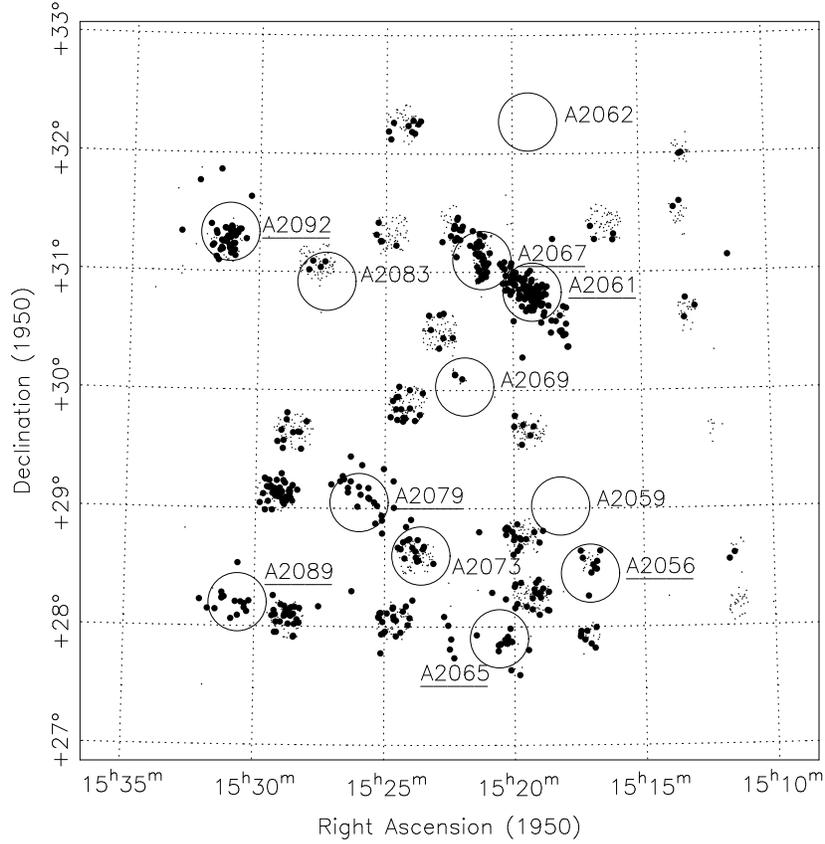}{4.3 in}{-90}{62}{62}{-234}{360}
\caption[]
{Location on the sky of galaxies with measured redshifts in the field
of the Corona Borealis supercluster.  The galaxies marked with large
dots are galaxies with redshifts placing them within the supercluster
($0.06 \le z \le 0.09$).  The twelve large circles, whose radii are
roughly $1 h^{-1}$ Mpc, mark the positions of the twelve Abell
clusters in the field.  The clusters whose names are underlined are
contained in the Corona Borealis supercluster.  The number of galaxies
successfully identified at all redshifts ranges from 10 to 42 for
fields observed when only half of the Norris Spectrograph's fibers
were usable because a large format 2048$^2$ CCD was not yet available
at Palomar, and from 59 to 87 for fields observed with the large
format 2048$^2$ CCD.  The precise locations of the observed fields are
given in Paper I.}
\label{figures:cb_radec}
\end{figure}

\begin{figure}
\plotfiddle{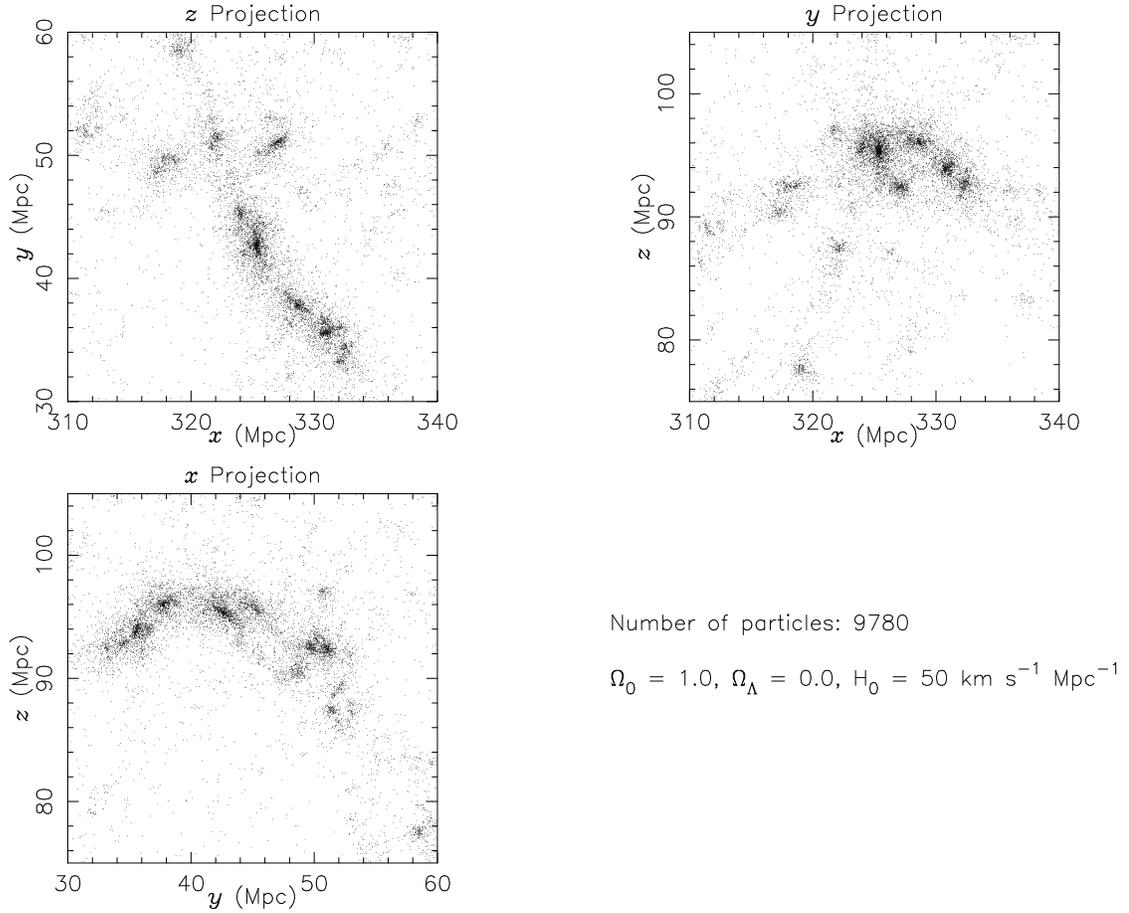}{4.3 in}{-90}{62}{62}{-234}{360}
\caption[]
{$x$, $y$, and $z$ projections of our simulated supercluster \#1.  Note
in the $z$ projection the relatively linear, chain-like structure.  To
simulate observing along the long axis of a cigar-shaped supercluster,
we rotate the supercluster in the $x$-$y$ plane by $45^\circ$ clockwise
and then observe along the $y$-axis.}
\label{figures:sim_sc}
\end{figure}

\begin{figure}
\plotfiddle{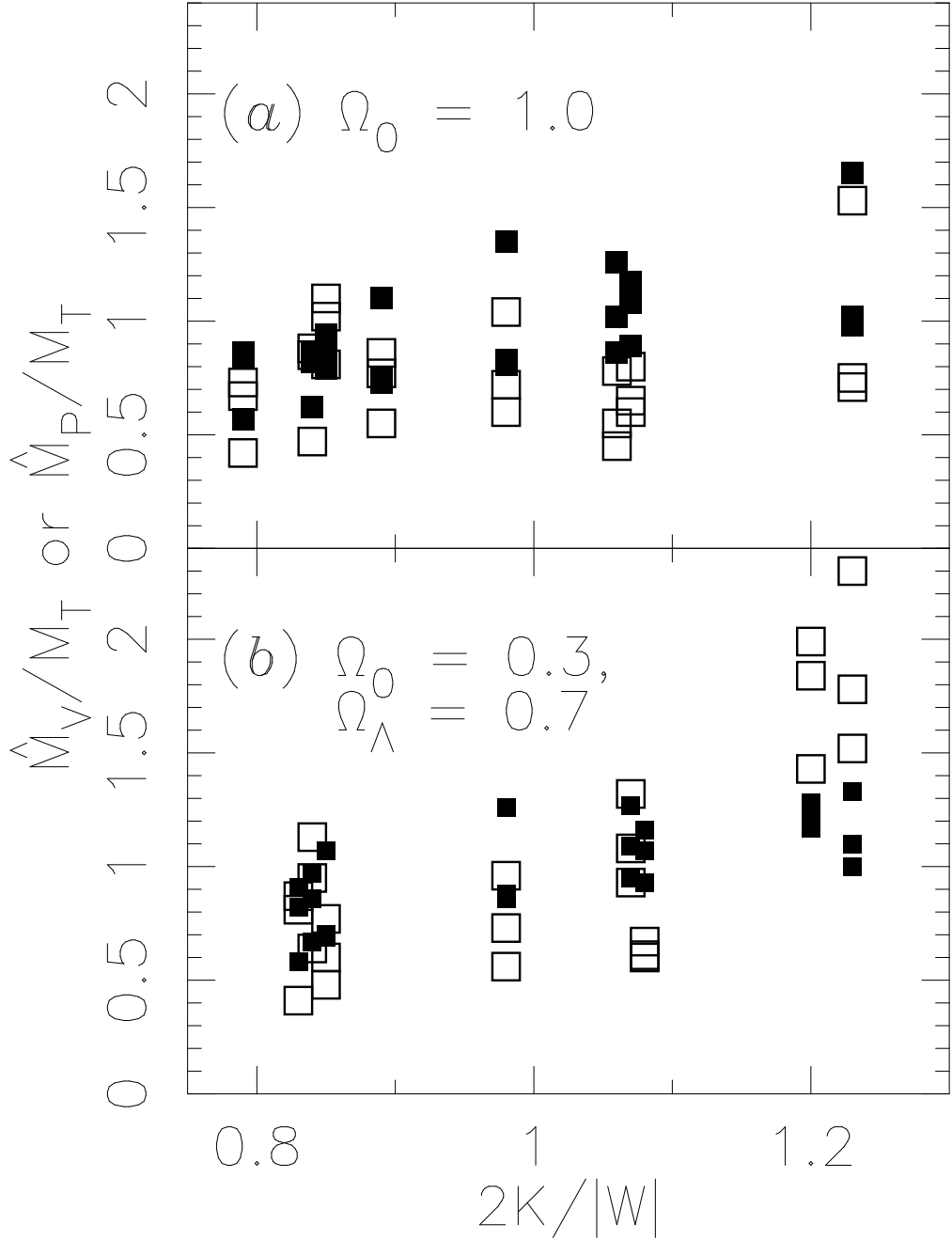}{6.5 in}{0}{85}{85}{-236}{-60}
\caption[]
{Ratio of the mass estimated with the virial mass estimator (filled
squares) or the projected mass estimator (unfilled squares) to the
true mass, as a function of $2K/|W|$ for our simulated superclusters.
Results from superclusters drawn from $\Omega_0 = 1$ simulations are
plotted in panel $(a)$, while results from superclusters from $\Omega_0
= 1$, $\Omega_\Lambda = 0.7$ simulations are plotted in panel $(b)$.
$\hat M_V/M_T$ is correlated with $2K/|W|$ as one would expect from
equation~(\ref{eq:2kw}).  The projected mass estimator can substantially
overestimate the mass of a supercluster with only a very small central
concentration of mass and a large amount of mass at large radii.}
\label{figures:mass_est}
\end{figure}

\begin{figure}
\plotfiddle{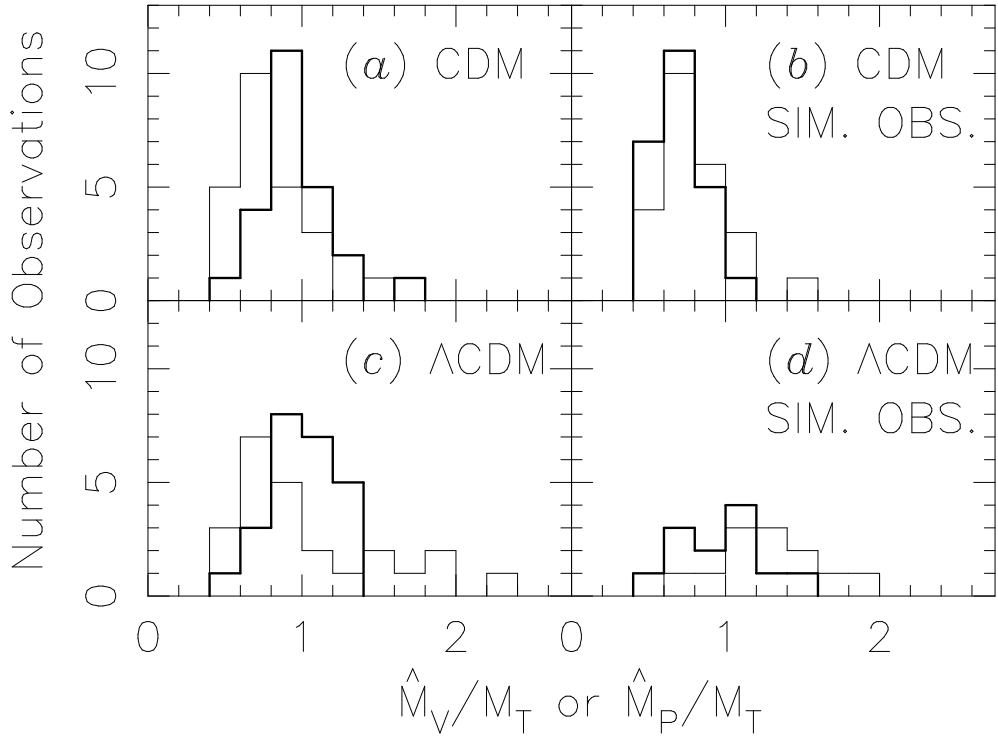}{4.3 in}{-90}{62}{62}{-234}{360}
\caption[] 
{Histograms of $\hat M_V/M_T$ (thick line) and $\hat M_P/M_T$ (thin line).
The histograms shown in panels $(a)$ and $(c)$ are based on mass
estimates using all of the simulation particles, while the histograms
shown in panels $(b)$ and $(d)$ are based on mass estimates from the
simulated observations.  The results for the superclusters drawn from
the $\Omega_0 = 1$ simulations are given in panels $(a)$ and $(b)$,
and the results for the superclusters drawn from the $\Omega_0 = 0.3$,
$\Omega_\Lambda = 0.7$ simulations are given in panels $(c)$ and
$(d)$.  Each observation is an observation along the $x$, $y$, or $z$
axis of one of the superclusters.  The results for only four
superclusters are plotted in panel $(d)$; see the note to Table 2.
The sparse and irregular sampling of the simulated observations causes
at most $\sim 30\%$ reductions in the accuracies of both the virial
mass estimator and the projected mass estimator.}
\label{figures:mass_hist}
\end{figure}

\begin{figure}
\plotfiddle{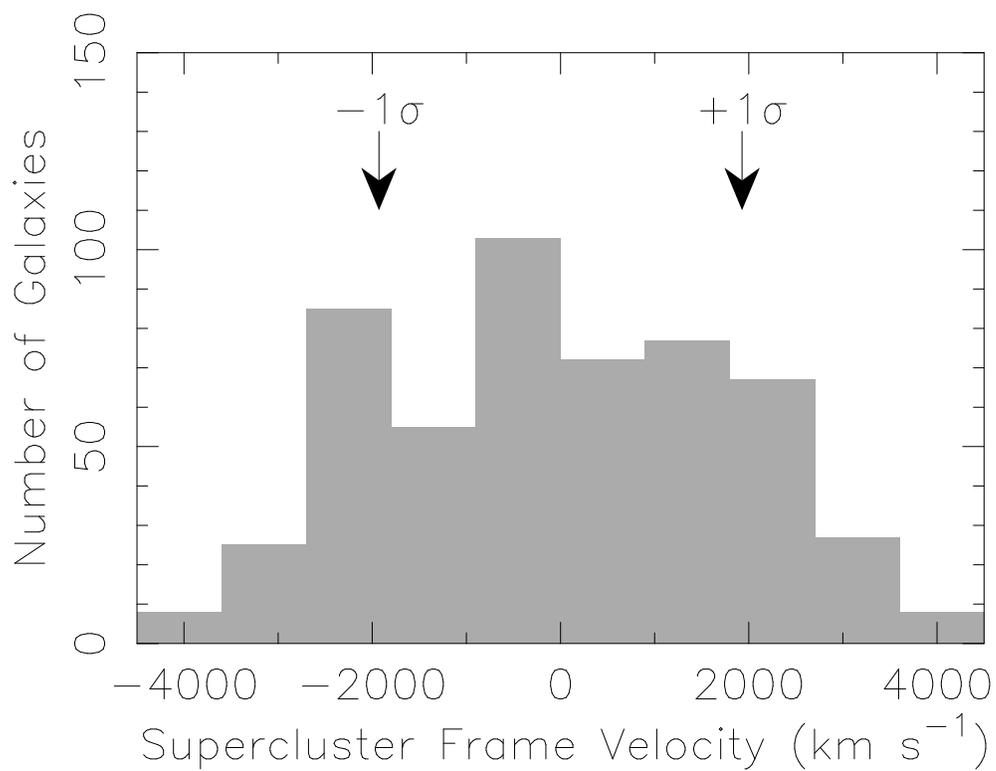}{4.3 in}{-90}{62}{62}{-234}{360}
\caption[]
{Line-of-sight velocity histogram for the Corona Borealis supercluster
in the supercluster frame.  We have marked $\pm 1\sigma$, where $\sigma$
is the velocity dispersion, with arrows.  The mean recession velocity
of the supercluster is $22420^{+149}_{-138}$ km s$^{-1}$, and
$\sigma = 1929^{+81}_{-67}$ km~s$^{-1}$.}
\label{figures:cor_bor_z}
\end{figure}

\begin{figure}
\plotfiddle{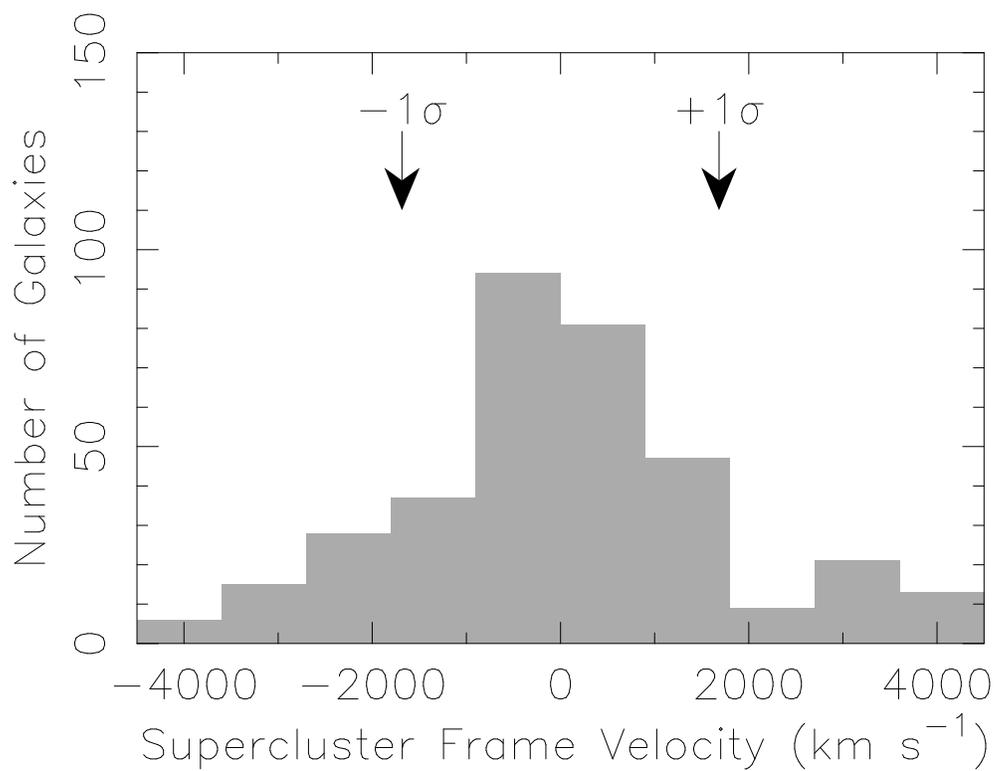}{4.3 in}{-90}{62}{62}{-234}{360}
\caption[]
{Line-of-sight velocity histogram for the A2069 supercluster
in the supercluster frame.  We have marked $\pm 1\sigma$, where $\sigma$
is the velocity dispersion, with arrows.  The mean recession velocity
of the supercluster is $34338^{+148}_{-140}$ km s$^{-1}$, and
$\sigma = 1684^{+145}_{-151}$ km~s$^{-1}$.}
\label{figures:a2069_z}
\end{figure}

\end{document}